\documentclass[conference]{IEEEtran}
\usepackage{graphics}
\usepackage{graphicx}
\usepackage{epsfig,amssymb}
\usepackage{epstopdf}
\usepackage{amsmath}
\usepackage{breqn}
\usepackage{times}
\usepackage{mathrsfs}
\usepackage{array}
\usepackage{amsmath, latexsym, amsfonts, amssymb}
\usepackage[mathscr]{eucal}
\usepackage{array, tabularx}
\usepackage[table]{xcolor}
\usepackage{multirow}
\usepackage{epsfig}

\usepackage{cite}
\usepackage[numbers,sort & compress]{natbib}

\usepackage{tikz}
\usepackage{url}
\usepackage{mathtools}
\usepackage{psfragx}
\usepackage{xcolor}

\newcommand{\paren}[1]{\left(#1\right)}
\newcommand{\sqparen}[1]{\left[#1\right]}
\newcommand{\brparen}[1]{\left\{#1\right\}}

\newcommand{\defeq}{\ensuremath{\triangleq}} 
\renewcommand{\vec}[1]{\ensuremath{\boldsymbol{#1}}} 

\newtheorem{theorem}{Theorem}
\newtheorem{lemma}{Lemma}
\newtheorem{definition}{Definition}

\IEEEoverridecommandlockouts

\begin{document}
\title{Wireless Energy Beamforming  using Signal Strength Feedback}
\author{
	\IEEEauthorblockN{Samith Abeywickrama}
	\IEEEauthorblockA{Department of Electronic and \\Telecommunication Engineering,\\University of Moratuwa, Sri Lanka.\\
		Email: samith@ent.mrt.ac.lk}\and	
	
	\IEEEauthorblockN{Tharaka Samarasinghe}
	\IEEEauthorblockA{Department of Electronic and \\Telecommunication Engineering,\\University of Moratuwa, Sri Lanka.\\
		Email: tharaka@ent.mrt.ac.lk}
	
	\and
	
	\IEEEauthorblockN{Chin Keong Ho}
	\IEEEauthorblockA{Institute for Infocomm Research,\\ Agency for Science, Technology and\\ Research (A-STAR), Singapore\\
		Email: hock@i2r.a-star.edu.sg}
	
	\vspace{-4cm}
	
	\thanks{More comprehensive work of this paper is published in IEEE Transactions on Signal Processing - \cite{8074795}}
}

\date{}
\bibliographystyle{ieeetr}
\maketitle
\vspace{-0.5cm}
\begin{abstract}
Multiple antenna techniques, that allow energy beamforming, have been
looked upon as a possible candidate for increasing the efficiency of the transfer process between the energy transmitter (ET) and the energy receiver (ER) in wireless energy transfer. This paper introduces a novel scheme that facilitates energy beamforming by utilizing Received Signal Strength Indicator (RSSI) values to estimate the channel. Firstly, in the training stage, the ET will transmit sequentially using each beamforming vector in a codebook, which is pre-defined using a Cramer-Rao lower bound analysis. The RSSI value corresponding to each beamforming vector  is fed back to the ET, and these values are used to estimate the channel through a maximum likelihood analysis. The results that are obtained are remarkably simple, requires minimal processing, and can be easily implemented. Also, the results are general and hold for all well known fading models. The paper also validates the analytical results numerically, as well as experimentally, and it is shown that the proposed
method achieves impressive results in wireless energy transfer. 
\end{abstract}

\section{Introduction}

Wireless energy transfer (WET) focuses on delivering energy to charge freely located devices, over the air interface, using   
Electromagnetic radiation in the radio frequency (RF) bands \cite{mimo_ck}. 
When it comes to RF signal enabled WET, increasing the efficiency of the energy transfer between the energy transmitter (ET) and the energy receiver (ER) is of paramount importance. Multiple antenna techniques that also enhance the range between the ET and the ERs have been looked upon as a possible solution to address this concern \cite{mimo_csi}. This paper proposes a novel approach that increases the efficiency of a WET system that utilizes multiple antennas to facilitate the energy transfer.

To this end, multiple antennas at the ET enable focusing the transmitted energy to the ERs via beamforming. However, the coherent addition of the signals transmitted from the ET at the ER depends on the availability of channel state information (CSI), which necessitates channel estimation at the ER in most cases. The estimation process involves analog to digital conversion and baseline processing which require significant energy \cite{channel_esti1,channel_esti2}. Under tight energy constraints and hardware limitations, such an estimation process may become infeasible at the ER. In this paper, we propose a more energy efficient method, that allows {\em almost} coherent addition of the signals transmitted from the ET at the ER. Moreover, this is a channel learning method that only requires feeding back Received Signal Strength Indicator (RSSI) values from the ER to the ET. In most receivers, the RSSI values are in fact already available, and no significant signal processing is needed to obtain them. It should be noted that the coherent addition of the signals transmitted from the ET at the ER depends directly on the phase of the channels, and it is interesting that our method focuses on estimating the required phase information by only using magnitude information about the channel.

Channel estimation in WET systems normally consists of two stages. The training stage, where feedback is obtained to estimate the channel, and the wireless power beamforming (WPB) stage, where the actual WET happens. Among the existing works that use multiple antennas and signal strength based channel estimation, \cite{rssi_work} proposes the following methodology. In the training stage, firstly, each antenna is individually activated, and then, antennas are pairwise activated, in order to obtain the RSSI values for each activation. Next, ignoring the noise, they utilize gathered RSSI values to estimate the channel. In \cite{one_bit_impl}, one-bit feedback algorithm \cite{one_bit} is used, and in the training stage, the receiver broadcasts a single bit to the transmitter indicating whether the current RSSI is higher or lower than in the previous, while the transmitter makes random phase perturbations based on the feedback of the receiver. 

Our proposed scheme is significantly different to \cite{rssi_work} and \cite{one_bit_impl}. We focus on a multiple-input-single-output (MISO) downlink consisting of two antennas at the ET and one antenna at the ER. 
The training stage consists of $N$ time slots. In each of these time slots, the ET will transmit using a beamforming vector from a pre-defined codebook. The ER feeds back the analog RSSI value corresponding to each beamforming vector, {\em i.e.,} the ET will receive $N$ RSSI feedback values. These $N$ feedback values are utilized to set the beamforming vector for the WPB stage. More precisely, feedback values are utilized to estimate the phase difference between the two channels between the ET and ER, and this difference is corrected when transmitting in the WPB stage, with the hope of adding the two signals coherently at the ER.

Our contributions and the paper organization can be summarized as follows. The system model and the methodology of obtaining feedback is explained in Section \ref{Section:System model}. In Section \ref{section:crlb}, we focus on defining the aforementioned pre-defined codebook. To this end, we employ a Cramer-Rao lower bound (CRLB) analysis, and define the codebook such that the estimator of the phase difference among the channels achieves the CRLB, which is the best performance that an unbiased estimator can achieve.
In Section \ref{Section:Channel Estimation}, we discuss how the feedback values can be utilized to set the beamforming vector for the WPB stage, through a maximum likelihood analysis. Our analysis takes noise into account unlike \cite{rssi_work}, and we present the no noise scenario as a special case. The results that we obtain are remarkably simple, requires minimal processing, and can be easily implemented at the ET. Also, the results are general such that they will hold for all well known fading models.  In Section \ref{section:Numerical Evaluations}, we validate our analytical results numerically. In Section \ref{Experimental Validation}, we go on to show that the proposed methodology can be in fact implemented on hardware. This is not common in the related works, and can be highlighted as another major contribution of this paper. To this end, we show that, our proposed method will achieve impressive results given how much power can be saved at the ER. It should be also noted that the proposed methodology can be used for any application of beamforming where processing capabilities of the receiver are limited. Section \ref{Section:Conclusions} concludes the paper.

\section{System Model and Problem Setup}\label{Section:System model}

We consider a MISO channel for WET. An ET consisting of $K$ antennas delivers energy to an ER consisting of a single antenna, over a wireless medium. For the clarity of the analysis, and due to the direct applicability to the experimental setup, it is assumed that $K=2$.\footnote{The results in this paper can be easily extended to $ K>2 $ scenario by using an approach similar to the one used in \cite{rssi_work} for the $K>2$ extension. Details are skipped due to space limitations.} For this setup, the received signal at the ER is given by 
\begin{eqnarray}
y =  \vec{h}^{\top} \vec{x} + z,\label{eq:ch model1}
\end{eqnarray}
where $\vec{h} = \sqparen {|h_{1}|e^{j\delta_{1}} \ \ |h_{2}|e^{j\delta_{2}}}^\top $ is the time-invariant complex random vector having an arbitrary distribution and representing the random channel gains between the ET and the ER, $\vec x $ is the $2$-by-$1$ vector representing the baseband transmit signal, and $z$ is complex random additive noise. We assume a quasi static block fading channel. The baseband transmit signal is defined as $\vec {x}= \vec {b}s $, where $s$ is the transmit symbol with unit power, and $\vec{b}$ is the $2$-by-$1$ beamforming vector. 

For this setup, the harvested energy at the ER is given by   
\begin{eqnarray}
Q = \xi \mathbb E \sqparen{\|y\|^{2}}, \label{eq:harvested energy}
\end{eqnarray}
where $\xi$ denotes the conversion efficiency of the energy harvester \cite{ck1}, and the expectation is performed over the random noise. It is not hard to see that for a given $\xi$, the energy transfer is maximized when $\|y\|$ is maximized, and this can be achieved by an optimal selection of $\vec{b}$. In practice, channel estimation is necessary to determine the optimal beamforming vector $\vec{b}$ that maximizes the energy transfer. However, we are particularly focusing on applications with tight energy constraints at the ER. Thus, such an estimation process may become infeasible as channel estimation involves analog to digital conversion and baseline processing, which require significant energy. Therefore, we focus on introducing a more energy efficient method of selecting the beamforming vector by only feeding back RSSI values from the ER to the ET. It should be noted that the feedback is a single analog value, and hence, no significant signal processing is required. In most receiver circuits, this RSSI value is in fact already available.   

The proposed scheme consists of a training stage and a WPB stage. The training stage consists of $ N $ time slots. 
In the $i$th training slot, where $i=1,\dots,N$, the ET uses beamforming vector $ \vec{b_{i}} $ for wireless beamforming, and the ER feeds back the analog signal strength, based on the measured RSSI, for the corresponding transmission. After completing the training stage, the ET will determine the beamforming vector $\vec{q}$ to be used for the WPB stage. 
The ER does not send any feedback in this stage, and typically, the WPB stage is longer than the training stage to reduce
the overhead incurred in the WPB. 

We define codebook $ \mathrm {\bf{B}} = \sqparen{\vec{b}_{1} \ 
\ldots \ \vec{b}_{N}} $ that includes $N$ beamforming vectors. Moreover, $ \vec{b}_{i} $ takes the form of $ \sqparen{1 \enspace e^{j\theta_{i}}}^\top$, where $\theta_{i}$ is the $i$th element in $\Theta$ ($i=1,\dots,N$), which is a predefined set that includes phase values between $ 0 $ and $ 2\pi $. For implementation convenience, the codebook is predetermined and does not depend on the signal strength feedback, but the WPB vector $ \vec q $ is designed based on all the signal strength feedback values. Further,  we shall employ estimation theory and the concept of the CRLB in order to define $\mathrm {\bf{B}}$. In the training stage, the pair of antennas at the ET is simultaneously activated for each element in $\mathrm {\bf{B}}$, and the corresponding RSSI value is fed-back through a wireless feedback channel. That is, we have $N$ RSSI values at the ET, and we focus on estimating a near optimal beamforming vector $ \vec q$ based on these RSSI feedback values, with a focus of combining the spatial signals from the ET coherently at the ER.  

Using \eqref{eq:ch model1}, and the proposed method of beamforming, the received signal that is related to the $i$th RSSI feedback value can be written as
\begin{eqnarray}
y_i =  |h_{1}|e^{j\delta_{1}}s + |h_{2}|e^{j\delta_{2}} e^{j\theta_{i}}s + z.  \label{eq:ch model2}
\end{eqnarray}
The corresponding $i$th instantaneous RSSI value can be expressed as 
\begin{alignat} {2}
\mathrm {R}_i 
&= \Big| |h_{1}|e^{j\delta_{1}}s + |h_{2}|e^{j(\delta_{2}+\theta_{i})}s \Big|^{2} +   w_i \nonumber\\
&=  \Big(|h_{1}|^{2} + |h_{2}|^{2} + 2|h_{1}||h_{2}|\cos\paren{\theta_{i}-\delta_{1}+\delta_{2}}\Big) + w_i \nonumber\\
&=\alpha + \beta \cos\paren{\theta_{i}+\phi} + w_i,
\label{eq:ch rssi}
\end{alignat} 
where $\alpha = |h_{1}|^{2} + |h_{2}|^{2}$, $\beta = 2|h_{1}||h_{2}|$, and $\phi= \delta_{2}-\delta_{1}$ (the phase difference between $h_{2}$ and $h_{1}$). We use $ w_i $ to represent the effect of noise on $ \mathrm R_{i} $. The noise term $ w_i $ includes all noise related to the measurement process, including noise in the channel, circuit, antenna matching network and rectifier. Since we are assuming a block fading model, $ \vec h $ can be considered to be unknown, but non varying (fixed) during the training stage and the subsequent beamforming. Therefore, the randomness in \eqref{eq:ch rssi} is caused only by the noise component $w_{i}$. For tractability, and without loss of generality, we assume $\vec{w}= \sqparen{w_1,\ldots, w_N}^\top $ to be an i.i.d. Gaussian random vector, having zero mean and variance $\sigma^2$. Also note that the Gaussian distribution leads to the worst-case CRLB performance for any estimation problem \cite{gau_assum}.  

From \eqref{eq:ch rssi}, it is easy to show that the RSSI value is maximized (leading to optimal energy transfer in the WPB) when $\theta_{i}= -\phi$, {\em i.e.}, the optimal beamforming vector $ \vec{b}_{\phi}= [1 \enspace  e^{-j\phi} ]^\top$. Hence, our goal is to estimate the phase difference of the two channels, and we denote the estimate using $\hat \phi$. Also from \eqref{eq:ch rssi}, the RSSI depends on two more unknown so-called nuisance parameters $ \alpha$  and $\beta$. Hence, the parameter vector is given by $ [\alpha \enspace \beta \enspace \phi]^{\top} $. Further, it can be shown that we need at least three RSSI values ($N \geqslant 3$) in order to estimate $ \phi $.\footnote{It can be shown that the CRLB is unbounded if $N < 3$ regardless of the choice of codebook. We do not provide formal proof details due to space limitations.}

To implement the proposed method in this paper, we should first define $\Theta$. In the next section, we define $\Theta$ by performing a CRLB analysis on the parameter vector. Then, $\Theta$ will be used to define the codebook $\mathrm {\bf{B}}$. In Section \ref{Section:Channel Estimation}, we discuss how the RSSI feedback values associated to the beamforming vectors in $\mathrm {\bf{B}}$ can be used to estimate $\phi$ through a maximum likelihood analysis.

\section{Cramer-Rao lower bound analysis } \label{section:crlb}

The CRLB is directly related to the accuracy of an estimation process. More precisely, the CRLB gives a lower bound on the variance of an unbiased estimator. To this end, suppose we wish to estimate the parameter vector  
$ \vec {\varphi} = [\alpha \enspace \beta \enspace \phi]^{\top} $. The unbiased estimator of $ \vec {\varphi} $ is denoted by $ \hat {\vec {\varphi}} = [\hat\alpha \enspace \hat\beta \enspace \hat\phi]^{\top} $, where $ \mathbb E \{\hat {\vec {\varphi}} \} = \vec {\varphi}$. The variance of the unbiased estimator $ \mathrm{var}( {\hat{\vec \varphi}}) $ is lower-bounded by the CRLB of $ \vec {\varphi} $ which is denoted by $ \mathrm{CRLB_{\vec {\varphi}}} $, \textit{i.e.},
$	\mathrm{var}(\hat{\vec \varphi}) \geqslant \mathrm{CRLB_{\vec {\varphi}}} $.
Moreover, $\mathrm{CRLB_{\vec {\varphi}}} $ is given by the inverse of $ \mathrm{FIM}_{\vec \varphi} $, which is the Fisher information matrix (FIM) of $\vec \varphi$.
Since no other unbiased estimator of $ \vec {\varphi} $ can achieve a variance smaller than the CRLB, the CRLB is the best performance that an unbiased estimator can achieve. Hence, we select $ \Theta $ such that the estimator achieves the CRLB, and hence, the variance is minimized. It should be also noted that the Gaussian distribution minimizes/maximizes the FIM/CRLB \cite{gau_assum2}. Therefore, due to the Gaussian assumption made on the noise power in \eqref{eq:ch rssi}, we are minimizing the largest or the worst case CRLB. 

Using \eqref{eq:ch rssi}, the $N$-by-$1$ vector representing $N$ RSSI observations can be written as
\begin{alignat}{2} 
\mathrm{\bf{R}}=  \vec x_{\varphi} +  \vec w,
\label{final_model}
\end{alignat}
where $\vec x_{\varphi} $ is a $N$-by-$1$ vector of which the $i$th element takes the form of $ \alpha + \beta \cos(\theta_{i} + \phi)$. Since $\vec x_{\varphi}$ is independent of $\vec w$, $\mathrm{\bf{R}}$ in \eqref{final_model} has a multivariate Gaussian distribution, {\em i.e.}, 
\begin{alignat*}{2} 
\mathrm{\bf{R}} \sim \mathcal{N} (\vec x_{\varphi}, \mathrm {\bf{C} }),
\end{alignat*}
where $ \mathrm {\bf{C} }= \sigma^2 \mathrm {\bf{I}_N} $ is the covariance matrix, and $\mathrm {\bf{I}_N}$ is the $N$-by-$N$ identity matrix. We will specifically focus on $ \phi $, which is the main parameter of interest, and derive the CRLB of its estimator. Then, we will find the set of values $\brparen{\theta_{i}}_{i=1}^N$ that will minimize the derived CRLB. The CRLB of $\phi$ is formally presented through the following lemma.
\begin{lemma} \label{Lemma: CRLB of phi}
The CRLB of parameter $ \phi $ is given by
\begin{alignat*}{1} 
\mathrm{CRLB}_{\phi} = \frac{\displaystyle \sigma^{2} \sum_{i=1}^{N-1} \sum_{j=i+1}^{N} \Big[\cos(\theta_{i}+\phi) - \cos(\theta_{j}+\phi) \Big]^{2} }  {\displaystyle \beta^{2}  \sum_{i=1}^{N-2}  \sum_{j=i+1}^{N-1}  \sum_{k=j+1}^{N} \Delta_{i,j,k}   },
\end{alignat*}
where
\begin{alignat*} {2}
\Delta_{i,j,k}=  \Big[\sin(\theta_{i}-\theta_{j}) + \sin(\theta_{j}-\theta_{k}) + \sin(\theta_{k}-\theta_{i}) \Big]^{2}. \end{alignat*}
\end{lemma}
We will only provide a sketch of the proof due to space limitations. The $i$th row of $\frac{\partial \vec x_{\varphi} }{\partial \vec \varphi}$ is given by 
$\sqparen{1 \quad \cos(\theta_{i}+\phi)  \quad -\beta \sin(\theta_{i}+\phi)}$, for $i=1,\ldots,N$. By using the FIM of a Gaussian random vector in \cite{crlb_book}, and using the fact that $\mathrm {\bf{C} }$ is independent of $\vec \varphi$, the FIM of $\mathrm{\bf{R}}$ can be written as $\mathrm{FIM}_{\varphi}(\mathrm{\bf{R}}) = \sqparen{ \frac{\partial \vec x_{\varphi} }{\partial \vec \varphi}   }^{\top} \mathrm {\bf{C} } \sqparen{ \frac{\partial \vec x_{\varphi} }{\partial \vec \varphi}   }$.
The CRLB of the $i$th element in $ \vec \varphi $ can be obtained by the $i$th diagonal element of the inverse FIM. Therefore, computing the third diagonal element of the inverse of $\mathrm{FIM}_{\varphi}(\mathrm{\bf{R}})$ completes the proof.

Since we want to find $\brparen{\theta_{i}}_{i=1}^N$ that will
minimize the derived CRLB for any given $\phi$, we average out
the effect of $\phi$ by considering the expectation over $\phi$. To this end, we assume $\phi$ to be uniformly distributed in $(0, 2\pi]$. This leads to the \textit{modified Cramer-Rao lower bound} (MCRLB) \cite{navigation}, and it is formally presented through the following lemma. The proof is skipped since its trivial.
\begin{lemma}
The MCRLB of parameter $ \phi $ is given by
\begin{alignat}{2} 
\mathrm{MCRLB}_{\phi} &= \mathbb E_{\phi} [\mathrm{CRLB}_{\vec \phi}]  \nonumber \\ 
&= \frac{\displaystyle \sigma^{2}\sum_{i=1}^{N-1} \displaystyle \sum_{j=i+1}^{N} \Big[1 - \cos(\theta_{i}-\theta_{j}) \Big] }  {\displaystyle \beta^{2} \displaystyle \sum_{i=1}^{N-2} \displaystyle \sum_{j=i+1}^{N-1} \displaystyle \sum_{k=j+1}^{N} \Delta_{i,j,k}   }. 
\label{crlb2}
\end{alignat}
\label{lemma2}
\end{lemma} 

Determining the $\brparen{\theta_{i}}_{i=1}^N$ analytically for a general case is not straightforward due to the complexity of \eqref{crlb2}. Therefore, we will first focus on the $N=3$ case, and derive  $\brparen{\theta_{1},\theta_{2},\theta_{3}}$ that minimizes the MCRLB. To this end, without any loss of generality, we assume  $\theta_{1}$ to be zero and $\theta_2$ and $\theta_3$ are set relative to $ \theta_{1} $. Then, we repeat the process for $N=4$. From these two derivations, we can observe a pattern in the $ \mathrm{MCRLB}_{\phi} $ minimizing $\theta_{i}$ values, and we define $\Theta$ by making use of this pattern. In Section \ref{section:Numerical Evaluations}, through numerical evaluations, we validate the selection of $\Theta$ for arbitrary values of $N$. 

\begin{lemma}
Let $\theta_1=0$. If $N=3$, $ \mathrm{MCRLB}_{\phi} $ is minimized when $\Theta=  \brparen{0 , 2\pi /3, 4\pi/3 }$, and the corresponding  minimum $\mathrm{MCRLB}_{\phi}$ is given by $\frac{2 \sigma^{2}}{3 \beta^{2}}$. If $N=4$, $ \mathrm{MCRLB}_{\phi} $ is minimized when $\Theta=  \brparen{0 , \pi/2,  \pi, 3\pi/2 }$, and the corresponding  minimum value of $\mathrm{MCRLB}_{\phi}$ is given by $\frac{2 \sigma^{2}}{4 \beta^{2}}$. 
\label{lemma3}
\end{lemma}
\begin{IEEEproof}
By differentiating \eqref{crlb2}  with respect to $ \theta_{2} $ and $ \theta_{3} $, respectively, and by setting $\theta_{1}=0$, we obtain two equations consisting of $ \theta_{2}$ and $ \theta_{3}$. Equating the two equations to zero and simultaneously solving them under the constraints $\theta_{2},\theta_{3} \in (0,2\pi]$ and  $\theta_{1} \neq \theta_{2} \neq \theta_{3} $ gives us $ \theta_{2}= 2\pi /3 $ and $ \theta_{3}= 4\pi/3 $. Evaluating the Hessian matrix at the stationary point $ \paren{0 , 2\pi /3, 4\pi/3 } $ shows that the stationary point is a minimum. Substituting $ \paren{0 , 2\pi /3, 4\pi/3 } $ in \eqref{crlb2} gives us $2\sigma^{2}/3 \beta^{2}$, which completes the proof for $N=3$. Following the same lines for the $N=4$ case completes the proof of the lemma.
\end{IEEEproof}

It is interesting to note that in both cases, the phase values in $\Theta$ are equally spaced over $ [0 \enspace 2\pi) $. For an example, when $N=3$, $ |\theta_{1} - \theta_{2}| = |\theta_{2} - \theta_{3}| = |\theta_{3} - \theta_{1}| = 2\pi/3$. When $N=4$, the phase difference between adjacent elements in the set turns out to be  $2\pi/4$. Also, by observing this pattern,we can expect  the minimum $ \mathrm{MCRLB}_{\phi} $ for arbitrary $N$ to take the form of $\frac{2 \sigma^{2}}{N \beta^{2}}$. To this end, we will define $\Theta$ for $N$ elements as follows.
\begin{definition}
$\Theta$ is a set of phase values between $0$ and $2\pi$, and it is defined to be $\Theta= \brparen{\theta_1, \ldots, \theta_N}$, where $\theta_i= \frac{2(i-1)\pi}{N}$ for $i \in \brparen{1,\ldots,N}$. 
\end{definition}

The intuition behind this definition is that getting RSSI values with the maximum spatial diversity provides us the best estimate. Using the phase values in $\Theta$, $N$ RSSI feedback values can be obtained. The next question is how these $N$ feedback values can be used to estimate the phase difference between the two channels. This, question is addressed in the next section.

%
%

\section{Estimation of Channel Phase Difference $ \phi $} \label{Section:Channel Estimation}

We will first look at a simplified scenario similar to \cite{rssi_work} by assuming that there is no noise. If there is no noise, we have $ \mathrm{R}_{i}= \alpha + \beta \cos{(\theta_{i}+\phi)} $, and we can consider $ N=3 $ and simply calculate $\phi$ by solving three simultaneous equations. The result is formally presented in the following theorem and this value of $\phi$ should intuitively give satisfactory results in low noise environments. The proof is skipped as it is trivial.  
\begin{theorem}
In a noiseless environment, the phase difference between the two channels is given by
\begin{alignat}{2}
	\hat\phi &= \tan^{-1}\paren{\frac{  \lambda_{1,3} \sin\paren{\frac{\theta_{1}+\theta_{2}}{2} } -  \lambda_{1,2}\sin\paren{\frac{\theta_{1}+\theta_{3}}{2} } }          
		{ \lambda_{1,2} \cos\paren{\frac{\theta_{1}+\theta_{3}}{2} } -  \lambda_{1,3} \cos\paren{\frac{\theta_{1}+\theta_{2}}{2} } 
		}	
	} 
	\raisetag{-.5em}
\end{alignat}
where $ \lambda_{i,j} = \mathrm {R}_{i} - \mathrm {R}_{j}$ and $i,j \in$ \{1,2,3\}. 
\end{theorem}
 
It should be noted that $ \phi $ has an ambiguity due to the use of $ \tan^{-1} $, and $ \phi $ can be either $ \phi $ or $ \phi - \pi $. The easiest way to resolve this ambiguity is by ascertaining two further RSSI feedback values from the ER for the two beamforming vectors $ [1 \enspace  e^{-j\phi} ]^\top  $ and $ [1 \enspace  e^{-j(\phi - \pi)} ]^\top  $ and picking the one that provides the better energy transfer. Also note that \cite{rssi_work} uses a similar approach, but it requires four more feedback values to resolve the ambiguity as the phase difference is given as a cosine inverse. 

Now, we will focus on a scenario with noise. Based on the assumption that the noise power is i.i.d. Gaussian, estimating $\phi$ becomes a classical parameter estimation problem. A maximum likelihood estimate of $\phi$ can be obtained by finding the value of $\phi$ that minimizes 
\begin{eqnarray*}
\mathrm E \defeq \sum_{i=1}^{N} \Big[\mathrm {R_{i}} - (\alpha + \beta \cos{(\theta_{i}+\phi)}) \Big]^{2}.
\end{eqnarray*}
Differentiating $\mathrm E$ with respect to  $\phi$, and setting it equal to zero gives us
\begin{equation}
	\sum_{i=1}^{N} \mathrm{ R_{i}}\sin{(\theta_{i}+\phi)} = \alpha \sum_{i=1}^{N} \sin{(\theta_{i}+\phi)} + \frac{\beta}{2} \sum_{i=1}^{N} \sin{[2(\theta_{i}+\phi)]}.
	\label{eq:estimation_phi}
\end{equation}
It is not hard to see that to estimate $\phi$, we have to first estimate $\alpha$ and $\beta$. These non-essential parameters are referred to as nuisance parameters \cite{nuisance}. However, due to the way we have defined $\Theta$, it is interesting to see that we can obtain an ML estimate of $\phi$ without estimating the nuisance parameters. These ideas are formally presented in the following theorem.  
\begin{theorem}\label{esti}
	For a sample of $ N $ i.i.d. RSSI observations,  $ \phi $ can be estimated by 
	\begin{gather}
	\hat\phi = \tan^{-1}\paren{\frac{\displaystyle  -\sum_{i=1}^{N} \mathrm{ R_{i}} \sin\theta_{i} }           
		{\displaystyle \sum_{i=1}^{N} \mathrm{ R_{i}} \cos\theta_{i}}	
	},
	\label{estimation_phi}
	\end{gather}
	where $ \theta_{i} = \frac{2(i-1)\pi}{N}$.
\end{theorem}
\begin{IEEEproof}
When $ \theta_{i} = {2(i-1)\pi}/{N}$, using series of trigonometric functions in \cite{ryzhik}, we have $\sum_{i=1}^{N} \sin(\theta_{i}+\phi)=\sum_{i=1}^{N} \sin{[2(\theta_{i}+\phi)]}=0 $.  Therefore, \eqref{eq:estimation_phi} can be simplified and written as $ \sum_{i=1}^{N} \mathrm{ R_{i}}\sin{(\theta_{i}+\phi)} = 0  $, which is independent of $\alpha$ and $\beta$. Expanding $\sin{(\theta_{i}+\phi)}$ allows us to obtain \eqref{estimation_phi}, which completes the proof.
\end{IEEEproof}

Here, $ \phi $ again has an ambiguity due to the use of $ \tan^{-1} $, and it  can also be resolved by ascertaining two more feedback values. Note that the result in Theorem \ref{esti} is easy to calculate, requires minimal processing, and can be easily implemented at the ET. Also, the result holds for all well known fading models. We should stress that this rather simple expression was possible due to the CRLB analysis performed in Section \ref{section:crlb} to define $\Theta$. In the next section, we will validate our results using numerical evaluations.  

\section{Numerical Evaluations} \label{section:Numerical Evaluations}

\begin{figure}[t] \vspace{-0.2cm}
	\centering{\includegraphics[scale=0.43]{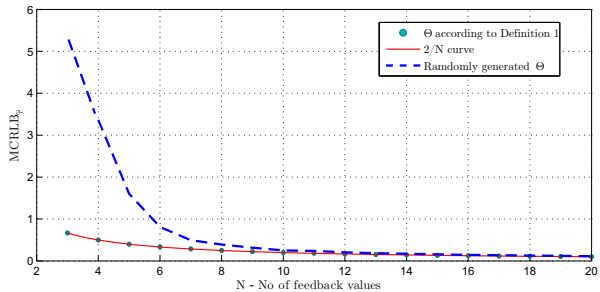}} 
	\caption{The behavior of the $ \mathrm{MCRLB}_{\phi} $ with N when  $ \beta = \sigma  = 1$.} 
	\label{Fig: CDF} \vspace{-0.3cm}
\end{figure}

In Lemma \ref{lemma3}, we have focused on $ \mathrm{MCRLB}_{\phi} $,  and we have given the formal proof for the minimum $ \mathrm{MCRLB}_{\phi} $, considering $ N=3 $ and $ N=4 $, respectively. Then, based on the pattern, we expected that the minimum $ \mathrm{MCRLB}_{\phi} = \frac{2 \sigma^{2}}{N \beta^{2}} $ for arbitrary values of $ N $. Fig. \ref{Fig: CDF} validates this result for arbitrary values of $ N $. For the numerical evaluation, we have set $\beta = \sigma =1 $, and we have calculated $\mathrm{MCRLB}_{\phi} $ according to Lemma \ref{lemma2}, while setting the phase values according to $\Theta$ in Definition 1. We can see that setting the phase values according to Definition 1 allows us to achieve the minimum MCRLB as the values lie on the $ 2/N $ curve. The figure also shows how the average $ \mathrm{MCRLB}_{\phi} $ behaves when the phase values in $ \Theta $ are chosen randomly, for a given $ N $. The average $ \mathrm{MCRLB}_{\phi} $ values always lie above the $ 2/N $ curve. Also, it can be seen that a $ \mathrm{MCRLB}_{\phi} $ value obtained by a randomly generated $ \Theta $ can be achieved by lower number of feedback values when $ \Theta $ is defined according to Definition 1. This is vital as we are dealing with a receiver with a tight energy constraint. Also, as expected, we can observe that when $ N $ increases, the lower bound on the variance of $ \hat{\phi} $ decreases. 

\begin{figure}[t]
	\centering{\includegraphics[scale=0.43]{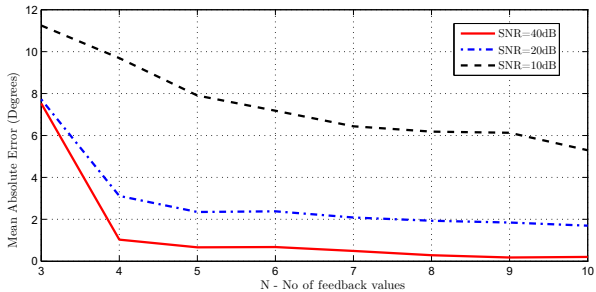}} 
	\caption{The behavior of the Mean Absolute Error (MAE) of $ \hat{\phi} $ for three different SNR values when $\beta = \sigma  = 1$.}	 
	\label{Fig: estimation} \vspace{-0.3cm}
\end{figure}

In Theorem \ref{esti}, we have presented an ML estimate of $ \phi $. Fig. \ref{Fig: estimation} illustrates the behavior of the phase estimation error with $ N $ for different SNR values. $\Theta$ is defined according to Definition 1. For the higher SNR values, error converges to zero rapidly than the lower SNR values. It is interesting to note that even when $ N=3 $, the phase error is not significantly large. Next, we will further validate our results experimentally.

\section{Experimental Validation}\label{Experimental Validation}


\begin{figure}[t]
	\centering{\includegraphics[scale=0.45]{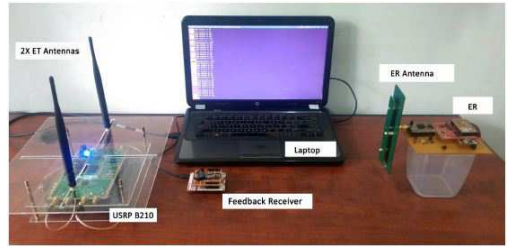}} 
	\caption{Experimental setup}	 
	\label{Fig: ET} 
\end{figure}

\begin{figure}[t]
	\centering{\includegraphics[scale=0.5]{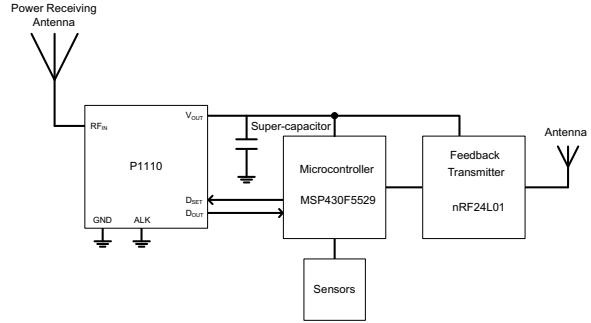}} 
	\caption{The hardware block diagram of ER.}	 
	\label{Fig: ER} \vspace{-0.1cm}
\end{figure}

\begin{figure}[t]
	\centering {\includegraphics[scale=0.44]{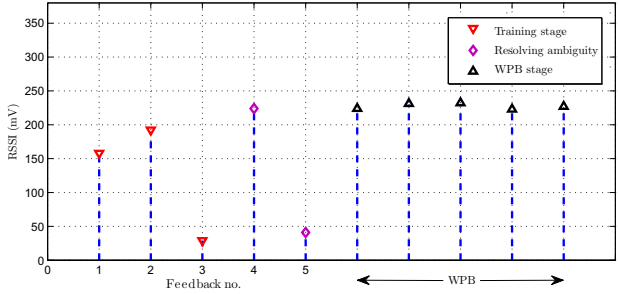}} 
	\caption{The RSSI values corresponding to each stage when   ET and  ER are 2 meters apart and $ N=3 $.}	 
	\label{Fig: rssi_3} \vspace{-0.2cm}
\end{figure}

The implementation of our ER is shown in Fig. \ref{Fig: ER}.
We use Powercast P1110 power-harvester, 
which has an operating band ranging from 902 to 928MHz. P1110 has an analog output ($ \mathrm {D_{OUT}} $), which provides an analog voltage level corresponding to the RSSI. 
As the storage device of our design, we use a low leakage
0.22F super-capacitor. 
The output of P1110 charges the super-capacitor and the super-capacitor powers the microcontroller, the feedback transmitter and the sensors. 
An Ultra-Low-Power MSP430F5529 microcontroller is used to read the RSSI values and transmit them via the feedback transmitter. 
When functioning, the microcontroller and the feedback transmitter are on sleep mode, and after each 500 ms interval, both wake up from sleep in order to read the RSSI and transmit it to the ET. NORDIC nRF24L01 single chip 2.4GHz transceiver has been used as feedback transmitter. 
When the ER operates in the active mode (reading RSSI values and transmitting), it consumes only 12.8 $ \mu $J/ms and it consumes negligible energy in sleep mode. 
The SDR used in our ET is USRP B210, which has $2  \times  2$ MIMO capability. CRYSTEC RF power amplifiers (CRBAMP 100-6000) are used to amplify the RF power output of the USRP B210. 
All the real-time signal processing tasks, channel phase difference ($ \phi $) estimation and setting beamforming vectors in both training and WPB stages were performed on a laptop using the GNU Radio framework. 
We use 915Mhz as the beamforming frequency. The same transceiver chip used in the ER, nRF24L01, is used as the feedback receiver at the ET side. For the experiment, the ET and the ER are 2 meters apart. Using this setup, for $ N=3 $, Fig. \ref{Fig: rssi_3} illustrates the training stage and the WPB stage, including ambiguity resolving, and we can see a clear gain by the proposed method.  

\begin{figure}[t]
	\centering {\includegraphics[scale=0.45]{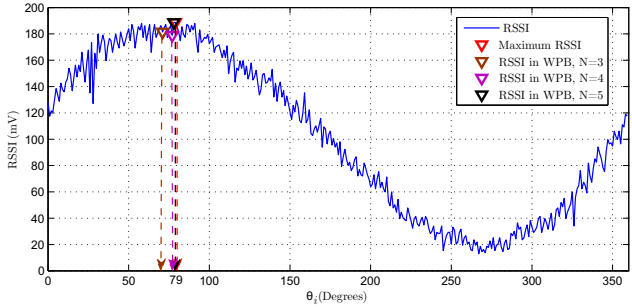}} 
	\caption{The RSSI values when $ \theta_{i} $ is changed from $ 0^{\circ} $ to $ 360^{\circ} $  with $ 1^{\circ} $ resolution.}	 
	\label{Fig: 0-360} \vspace{-0.5cm}
\end{figure} 

\begin{table} 
	\caption {Experimental results} \vspace{0.4cm}
	\rowcolors{1}{}{lightgray}
	\centering \begin{tabular}{  p{1.5cm}  p{1.5cm}  p{1.7cm}  }
		\hline \centering  N & \centering $ \hat \theta $ & Error $ | \hat \theta -79^{\circ}| $ \\
		\hline \centering  3 & \centering $ 71^{\circ} $ & $ \qquad 8^{\circ} $ \\
		\centering  4 & \centering $ 77^{\circ} $  & $\qquad 2^{\circ} $ \\
		\centering  5 & \centering $ 78^{\circ} $ & $\qquad 1^{\circ} $ \\ 
		\centering  6 & \centering $ 76^{\circ} $ & $\qquad 3^{\circ} $ \\                        	
	\end{tabular} 
	
\end{table} 

Then, we focused on validating the result on phase estimation. For this, we changed $ \theta_{i} $ from $ 0 $ to $ 360 $ degrees with $ 1^{\circ} $ resolution, and collected all respective RSSI values (see Fig. \ref{Fig: 0-360}). Since it was not practical to collect all the 360 RSSI values using the harvested energy via the feedback transmitter, we used a wired feedback for this experiment. 
Fig. \ref{Fig: 0-360} shows that the maximum RSSI occurs when $ \theta_{i} = 79^{\circ}$. Therefore, the maximum energy transfer happens at that point. 
Using the same set of values, we estimated $\hat \phi $ ($\Theta$ defined according to Definition 1) for $ N=3 $, $ N=4 $, $ N=5 $ and $ N=6 $, respectively. The results are tabulated in Table I. It is not hard to see that the errors are significantly small, and they are consistent with the numerical evaluations as well. Further, by using our proposed scheme, and based on the assumption that the conversion efficiency of the power-harvester is fixed, we can extend the range of the ER by 52\% on average. This has been calculated based on the experimental results considering free space loss.  

\vspace{-0.2cm}

\section{Conclusions}\label{Section:Conclusions}

This paper has proposed a new channel estimation approach to be used in 
a multiple antenna WET system. 
The ET will transmit using beamforming vectors from a codebook, which has been pre-defined using a Cramer-Rao lower bound analysis. RSSI value corresponding to each beamforming vector is fed back to the ET, and these values have been utilized to estimate the channel through a maximum likelihood analysis. The results that have been obtained are simple, requires minimal processing, and can be easily implemented. The paper has also validated the analytical results numerically, as well as experimentally.
It has been shown that the results in the paper are more appealing as compared to existing channel estimation methods in WET, especially when there is tight energy constraints and hardware limitations at the ER. It is also important to point out that although channel estimation for multiple antenna WET systems has been considered in this paper, the proposed methodology can be used for any application of beamforming where processing capabilities of the receiver are limited. Further analytical and experimental results related to the proposed methodology will be presented in future work. 

\bibliography{bibfile}
\end{document}